\documentclass[sigconf]{acmart}
\usepackage{enumitem}

\usepackage{listings}
\usepackage{lipsum}

\definecolor{TODOcolor}{HTML}{FF0000}
\definecolor{AddContentcolor}{HTML}{0000CC}
\definecolor{REVISEcolor}{HTML}{FF0000}

\newcommand{\revise}[0]{}
\definecolor{personalcolor}{HTML}{516AA1}
\definecolor{sociocolor}{HTML}{E88C3B}
\definecolor{physicalcolor}{HTML}{4CB77F}
\newcommand{\person}[1]{\textcolor{personalcolor}{#1}}
\newcommand{\socio}[1]{\textcolor{sociocolor}{#1}}
\newcommand{\phys}[1]{\textcolor{physicalcolor}{#1}}

\AtBeginDocument{%
  }

\copyrightyear{2026}
\acmYear{2026}
\setcopyright{cc}
\setcctype{by}
\acmConference[UIST '26]{The 39th Annual ACM Symposium on User Interface Software and Technology}{November 02--05, 2026}{Detroit, MI, USA}
\acmBooktitle{The 39th Annual ACM Symposium on User Interface Software and Technology (UIST '26), November 02--05, 2026, Detroit, MI, USA}
\acmDOI{10.1145/3830398.3830560}
\acmISBN{979-8-4007-2856-3/2026/11}

\begin{document}

\title{SiMUSation: An Interactive Visitor Experience Simulation Framework to Support Museum Exhibition Design}


\author{Huanchen Wang}
\orcid{0000-0001-9339-1941}
\affiliation{
  \institution{Southern University of Science and Technology}
  \city{Shenzhen}
  \state{Guangdong}
  \country{China}
}
\affiliation{
  \institution{City University of Hong Kong}
  \city{Hong Kong}
  \country{China}
}
\email{wanghc2022@mail.sustech.edu.cn}

\author{Qiuming Chen}
\orcid{0009-0009-2859-2822}
\affiliation{
  \institution{Southern University of Science and Technology}
  \city{Shenzhen}
  \state{Guangdong}
  \country{China}
}
\email{12432735@mail.sustech.edu.cn}

\author{Zhonghao Ji}
\orcid{0009-0000-1788-0831}
\affiliation{
  \institution{Southern University of Science and Technology}
  \city{Shenzhen}
  \state{Guangdong}
  \country{China}
}
\email{12432740@mail.sustech.edu.cn}

\author{Ruqi Sun}
\orcid{0009-0004-2662-0379}
\affiliation{
  \institution{Southern University of Science and Technology}
  \city{Shenzhen}
  \state{Guangdong}
  \country{China}
}
\email{sunrq2024@mail.sustech.edu.cn}

\author{Zhichao Lu}
\orcid{0000-0002-4618-3573}
\affiliation{
  \institution{City University of Hong Kong}
  \city{Hong Kong}
  \country{China}
}
\email{zhichao.lu@cityu.edu.hk}

\author{Yuxin Ma}
\authornote{Corresponding author.}
\orcid{0000-0003-0484-668X}
\affiliation{
  \institution{Southern University of Science and Technology}
  \city{Shenzhen}
  \state{Guangdong}
  \country{China}
}
\email{mayx@sustech.edu.cn}


\begin{abstract}
Understanding how diverse audiences engage with narratives and content is central to exhibition design, yet designers often rely on intuition. Existing experience evaluation methods are typically retrospective, costly, and offer limited access to visitors' internal states, hindering early-stage iterative refinement.
Rather than relying only on post-implementation evaluation with real visitors, we explore LLM-driven persona simulation as a reference for early-stage design.
Following this idea, we present SiMUSation, an interactive framework designed to support early-stage exhibition design. SiMUSation models diverse visitor personas and simulates their exhibition experiences through a dual-layer representation that couples observable behaviors, such as movement and gaze, with corresponding internal responses, such as confusion and narrative engagement. Designers can steer simulations, inspect feedback from simulated visits, and iteratively revise layouts, content, and narrative flow to further examine how changes reshape visitor experience.
We implemented a prototype and evaluated it through a user study (N=12), showing that SiMUSation provides insights for reflection and refinement in early-stage exhibition design.
Our findings further highlight the potential of persona-driven simulation to support audience-informed evaluation and iterative decision-making across design tasks.
\end{abstract}



\setlength{\abovecaptionskip}{0.18cm} 
\begin{teaserfigure}
\vspace{-2mm}
    \includegraphics[width=\textwidth]{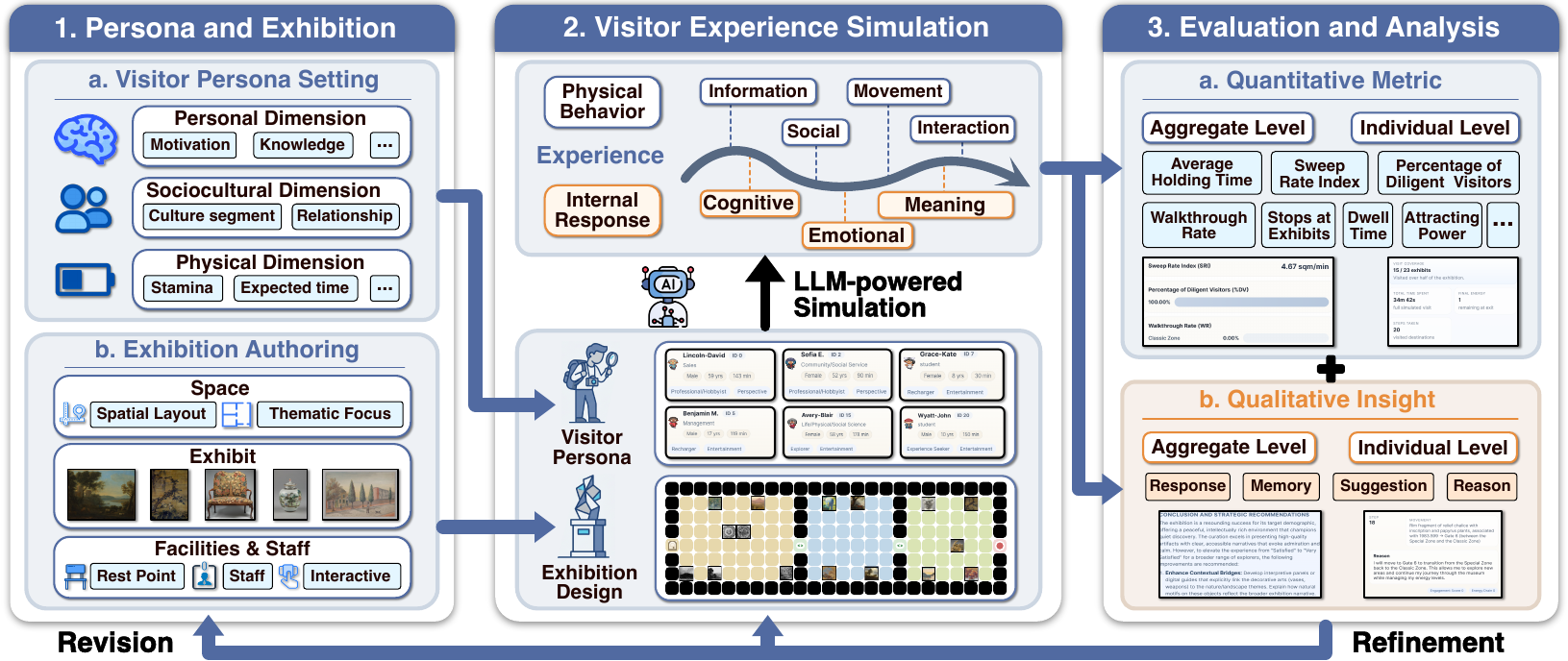}
  \caption{The SiMUSation framework consists of: 1) Persona \& Exhibition, supporting the diverse and plausible visitor personas building and the interactive authoring of exhibition space and narrative; 2) Visitor Experience Simulation, which simulates dynamic experiences from the defined personas and exhibition environment, with feedback represented as physical behaviors and internal responses; and 3) Evaluation \& Analysis, providing multi-faceted analysis to inform early-stage design iteration.}
  \Description{This figure illustrates the overall SiMUSation framework, which consists of three stages: Persona and Exhibition, Visitor Experience Simulation, and Evaluation and Analysis. The first stage supports the construction of diverse and plausible visitor personas together with the interactive authoring of exhibition space and narrative. The second stage simulates dynamic visitor experiences based on the defined personas and exhibition environment, with feedback expressed through physical behaviors and internal responses. The third stage provides multi-faceted evaluation results to support analysis and early-stage design iteration.}
  \label{fig:teaser}
\end{teaserfigure}
\maketitle

\section{Introduction}
Understanding how a museum exhibition’s narrative is perceived, interpreted, and reconstructed by its audience is central to exhibition design and evaluation~\cite{rogers2012audience, george2015curator, falk2016identity}. Yet designers still rely largely on professional intuition, making design outcomes susceptible to individual perspectives and biases~\cite{vergo1997museology, falk2016museum}. Because the effects of design decisions often remain unclear until implementation or public opening, evaluation typically depends on costly observation and visitor feedback collected after the fact~\cite{falk2016museum, hillier2006space, wineman2010space, serrell1997paying}. These constraints limit early-stage assessment and iterative refinement.

Prior work has explored digital approaches for capturing and analyzing personalized visitor experiences in museums~\cite{ivanov2025digitalex, wieland2026evaluate, almeshari2021guide}, but these approaches remain limited: \revise{behavioral data cannot fully account for visitors’ internal experiences~\cite{lanir2017visualizing, yoshimura2014analysis, ferrato2022behavior, emerson2020behavior}, and user studies are difficult to scale and often lack representativeness~\cite{roussou2013questionnaires, rey2020questionnaires}.}
Recent advances in large language models (LLMs), especially in LLM-based role-playing~\cite{shanahan2023role, chen2024llmrole}, suggest a promising alternative. By constructing coherent and diverse personas, LLMs can generate complex human behaviors and responses across contexts, as demonstrated in domains such as social simulation, games, and writing~\cite{wang2024voyager, park2023simulation, lim2025gamepersona, han2024writing}.
However, both traditional digital techniques and recent LLM-based museum works have primarily focused on visitor-facing support, including navigation and guidance~\cite{lee2024explore, almeshari2021guide}, VR/AR-based interactive experiences~\cite{wieland2026evaluate, meinichk2022virtual, xu2024AR}, and LLM-based role and interaction design~\cite{bu2026reframing, nielsen2025npc, jiang2025avatar, chen2025roles, su2025simviews}.
In contrast, support for curators and exhibition designers during early-stage design remains underexplored.

\revise{Motivated by these gaps and informed by a formative study (N=7), we present SiMUSation, an interactive framework for early-stage museum exhibition design that organizes the workflow as an authoring-simulating-iterating loop. SiMUSation integrates LLM-powered visitor simulation into the exhibition design workflow, linking physical behavior with internal responses to provide interpretable representations of visitor experience feedback and support reflection and design iteration.}
We implemented SiMUSation as a prototype and evaluated its usability and effectiveness through a user study (N=12). 
\revise{The results suggest that SiMUSation can provide actionable insights into visitor experience and support design refinement by helping designers identify where issues arise through behavioral traces and why through simulated internal responses.} 
Based on these findings, we discuss implications for LLM-powered agent systems in user-related design tasks. Our contributions are:

\begin{itemize}[leftmargin=*]
\item \revise{We conducted a formative study that identified challenges in early-stage exhibition design workflows and derived design goals.}
\item \revise{We present SiMUSation, an interactive framework that embeds LLM-powered visitor experience simulation into the exhibition design workflow for iterative refinement.}
\item \revise{We developed an interactive system and evaluated the framework's usability and effectiveness in supporting design refinement through a user study.}
\end{itemize}

\section{Background and Related Work}
\subsection{Visitor Experience in Museums}
In museums, visitors are not merely viewers of displayed objects but active participants in meaning-making and experience construction~\cite{silverman1995meaning, diamantopolou2012meaning, falk2016museum}. Accordingly, museums are widely understood as places of experience, where people come not only to learn, but also to enjoy leisure and socialize with others~\cite{King2023MEUX, silverman1995meaning, falk2016museum}. 
Designing engaging visitor experiences has therefore become a central concern in contemporary museum practice, as museums seek to create experiences that are meaningful, inclusive, and memorable and to sustain public engagement across diverse audiences~\cite{jones2015audience, seagram1993audience}.

Substantial research has examined how museum experiences are formed and interpreted~\cite{hooper2006studying, falk2016museum, packer2016conceptualizing, kirchberg2012experience}. The visitor experience can be modeled as multi-phased and context-dependent, spanning before, during, and after the visit~\cite{falk2016museum}. It thus frames museum visiting as a dynamic and situated experience rather than a purely observational activity. Prior studies further suggest that such experiences are shaped through the interaction of personal, social, and physical contexts, together with visitors' feelings and memories~\cite{falk2016identity, falk2016museum, packer2016conceptualizing}. At the same time, this complexity makes it difficult to capture and assess the visitor experience holistically, which in turn limits efforts to evaluate and refine exhibition experiences~\cite{falk2016museum, King2023MEUX, hein2002learning}.

Digital methods have been increasingly applied to analyze museum visitor experience~\cite{grack2015data, hornecker2022human, serrell1997paying}. Prior work often uses sensing technologies, such as Bluetooth~\cite{yoshimura2014analysis}, cameras~\cite{ferrato2022behavior, hong2021camera}, and complex multi-device systems~\cite{lanir2017visualizing, rochi2004system, ardissono2012system}, to collect physical behavioral data and identify patterns in related visitor measures that support evidence-based assessment. However, visitor experience is inherently complex, and behavioral data alone cannot fully capture visitors' interpretations, emotions, and reflections~\cite{addis2024museum}. To access these experiential dimensions more directly, researchers often use surveys and interviews to examine what visitors learn, think, and feel~\cite{roussou2013questionnaires, rey2020questionnaires, othman2011questionnaires, bollo2005analysis}. Yet these methods are difficult to scale and labor-intensive to conduct and analyze~\cite{grack2015data}. Therefore, these limitations hinder institutions and designers from systematically analyzing visitor experience for exhibition evaluation and design improvement.

Digital technologies have been widely used to enhance museum experience~\cite{shah2018enhance, ivanov2025digitalex}. Prior work has supported visitor navigation, guidance, and information access through positioning and mobile systems~\cite{almeshari2021guide,hage2010mobile, rubino2013musa} and robotic guides~\cite{hellou2022robotic,yoshinori2007robotic}. Related studies have also explored tangible and embodied interactions, enabling visitors to engage with exhibitions~\cite{wieland2026evaluate}. Building on these approaches, AR and VR link physical and virtual exhibits and spaces to support immersive exploration~\cite{xu2024AR, cheng2024AR, shehade2020VR, schofield2018VR}. This line of work further extends to virtual museums, which examine how exhibitions can be designed and experienced in fully digital environments, including screen-based and panoramic settings~\cite{meinichk2022virtual, li2025virtual, chalmers2024virtual}.
Building on prior work, we shift attention from visitors themselves to curators and designers, positioning visitor experience as a lens integrated into the exhibition design workflow at an early stage.

\vspace{-1.5mm}\subsection{LLMs for Museum Exhibition}
Recent advances in LLMs, including awareness, emotion, and personality, have enabled more human-like responses and stronger adherence to personas and roles ~\cite{shanahan2023role, scherrer2023LLM, huang2024humanity, huang2024emotional, li2024LLM}. 
\revise{Building on this progress, recent studies have developed LLM role-playing approaches that use persona descriptions and role profiles to shape agent behavior in different tasks and interaction contexts, with some approaches representing both internal thoughts and physical actions in settings~\cite{chen2024llmrole, tseng2024llmpersona, tang2025roleplaying, wang2025thoughtAction}.
As a result, LLMs are increasingly used in role-playing applications across domains such as game design~\cite{lim2025gamepersona, wang2024voyager}, writing~\cite{han2024writing,agrawal2023comic, ran2025bookworld}, programming~\cite{hong2024metagpt, zhu2026adcoder}, business~\cite{weng2025business, mccloskey2024natural}, and social simulation~\cite{park2023simulation, ale2023using}, where they can represent diverse beliefs, perspectives, and tendencies to generate personalized and situated content and decisions for various user-centered task-solving~\cite{chen2024llmrole, Ha2024persona, li2026simulation4design}.}
In parallel, growing human-AI co-creation research has examined how to understand and interact with LLM-powered personas and roles to support several creative tasks~\cite{Ha2024persona}, including ideation~\cite{choi2025creating, liu2025ideation}, graphic design~\cite{shin2025poster}, writing~\cite{talaei2025writing, park2026writing}, and game design~\cite{jo2026gaming}. These developments suggest that visitor experience simulation as a reference may also support early-stage museum exhibition design.

In museum and exhibition contexts, recent LLM-based work has primarily focused on enhancing visitor-facing personalized interaction and guidance~\cite{bu2026reframing}. One main direction uses LLM-powered guides, such as docent agents and robotic guides, to provide explanation and exploration support~\cite{almeshari2021guide, nielsen2025npc}. Another direction introduces interactive exhibition-related artifacts and characters, such as historical figures, to enrich engagement and interpretation~\cite{wang2024virtuawander, jiang2025avatar}. Related studies also explore the role of LLM and how LLM with multiple perspectives can help visitors sequence content, understand knowledge, and engage with the exhibition in more adaptive ways~\cite{su2025simviews, li2025eyesee, chen2025roles}. However, little work has investigated a more comprehensive framework for visitor experience simulation and examined how such simulation can directly support exhibition evaluation and design iteration. This gap motivates our framework, which enables rich visitor experience simulation and experience-driven, iterative editing to support a closed design loop.

\section{Formative Study}
To identify challenges in early-stage exhibition design, we conducted a formative study with 7 participants (aged 20-50 years, M=30.86; 3 females and 4 males) recruited via campus forums and promotional referrals, including museum-exhibition-related domain practitioners, researchers, and students~(\autoref{table:formative participants}). Each participant joined in a 90-minute Zoom session and received 100 CNY (approximately USD 14) in compensation. The study aimed to examine their current design practices and how they incorporate visitor experience into design thinking. In each session, participants first completed a brief painting exhibition design task based on a floor plan, during which they developed a narrative, marked intended visitor routes, and identified areas of uncertainty~(Appx.~\ref{A:brief task}). This was followed by a semi-structured interview about their experiences and challenges in conventional design workflows spanning the stages of planning, evaluating, and refining exhibition designs,  as well as their expectations and preferences for future support.

\vspace{-1.5mm}\subsection{Findings}

\subsubsection{F1: Difficulty in Anticipating and Considering Diverse Visitor Experiences}
During early-stage exhibition planning, all participants emphasized the importance of adopting a visitor perspective by considering movement paths, narrative comprehension, and different audience groups. At the same time, they noted that analyzing visitor experience requires accounting for many nuanced and intersecting factors, including prior knowledge (P1, P2, P4, P6), interest (P3), sociocultural background (P3, P6, P7), motivation (P3, P4, P5, P6), and physical constraints such as stamina (P3, P7) and expected visit duration (P5). The complexity of these factors made it difficult for participants to anticipate diverse responses and translate them into design decisions. 
\revise{Several participants (P1-P5) further noted that, without concrete ways to examine how different visitor groups might experience an exhibition, they often relied on their own professional experience, which could introduce personal biases into early-stage exhibition design planning.}

\vspace{-1mm}\subsubsection{F2: Lack of Real-time and Multi-faceted Evaluation of Experience}
Most participants also emphasized that effective evaluation and design refinement should be grounded in visitor experience. However, they noted that existing evaluation methods are mostly retrospective, which are conducted only after an exhibition is installed or opened (P3, P7). Without feedback before deployment, identifying and correcting design flaws can become costly. 
All participants argued that current approaches do not adequately capture the dynamic and complex nature of visitor experience. 
\revise{Several participants (P1, P3, P6) further noted that although observable behaviors such as movement paths and dwell times can be recorded, they provide limited access to visitors' internal responses, including narrative understanding and emotional engagement.}
P6 and P7 both stressed that visitor experience should not be assessed by a single standard, since different audiences vary in their visiting contexts.
Therefore, participants expressed a strong need for evaluation methods that can provide timely, multi-faceted, and audience-sensitive feedback to support visitor understanding and evaluation.

\vspace{-1mm}\subsubsection{F3: Limitations of Static Methods for Exhibition Authoring and Iteration}
Many participants described revision and refinement as difficult parts of early-stage exhibition design. They often rely on static formats such as document outlines (P1, P6) and rough sketches (P3, P4, P6, P7), which made changes cumbersome and fragmented. 
These workflows also provided limited support for informed revision, making iteration largely trial and error. Several participants therefore wanted more practical feedback to support revision, including both quantitative summaries and qualitative comments (P1, P5). 
\revise{Participants (P1-P4, P6-P7) also noted that, while not ideal, 2D canvas authoring was sufficient for early-stage design.}
When discussing potential authoring support, participants consistently emphasized the importance of preserving creative control. Rather than expecting tools to generate finalized solutions, they preferred tools that offer references and suggestions while leaving design decisions to the designer.

\vspace{-2mm}\subsection{Design Goals}
To address these challenges identified in the findings, we derived three design goals for an LLM-powered framework that simulates complex and diverse visitor experiences to support early-stage exhibition design, refinement, and revision:

\vspace{-1mm}\subsubsection{DG1: Support Reliable and Diverse Visitor Personas Building}
This goal supports the construction of heterogeneous visitor personas, helping designers move beyond subjective assumptions (F1). The framework allows users to define intersecting and multidimensional factors grounded in prior literature and our formative study findings, enabling early-stage design to better account for diverse and plausible visitor experiences.

\vspace{-1mm}\subsubsection{DG2: Provide Dynamic Simulation and Multi-Granular Feedback}
To address the retrospective conventional evaluation and the "black box" nature of visitor experience (F2), the framework simulates exhibition visits based on diverse visitor personas (F1) and provides a visual and interpretable visit process by connecting observable behaviors with internal responses over time. In addition, the framework provides multi-granular feedback, combining quantitative metrics with qualitative insights at both the aggregate and individual levels, to support timely and multi-faceted analysis. Such feedback can help designers move beyond intuition and identify potential design issues for refinement and revision (F2, F3).

\vspace{-1mm}\subsubsection{DG3: Facilitate \revise{Insight-Driven} Iteration via Interactive Authoring}
To facilitate more flexible, experience-based iteration during early-stage design (F2, F3), the framework seamlessly integrates simulation feedback into an interactive authoring environment. Rather than automated solutions, it provides suggestions and clues as reference, allowing designers to iteratively refine layout and narrative plans while maintaining creative agency (F3).

\begin{figure*}[!htbp]
\centering
\includegraphics[width=\textwidth]{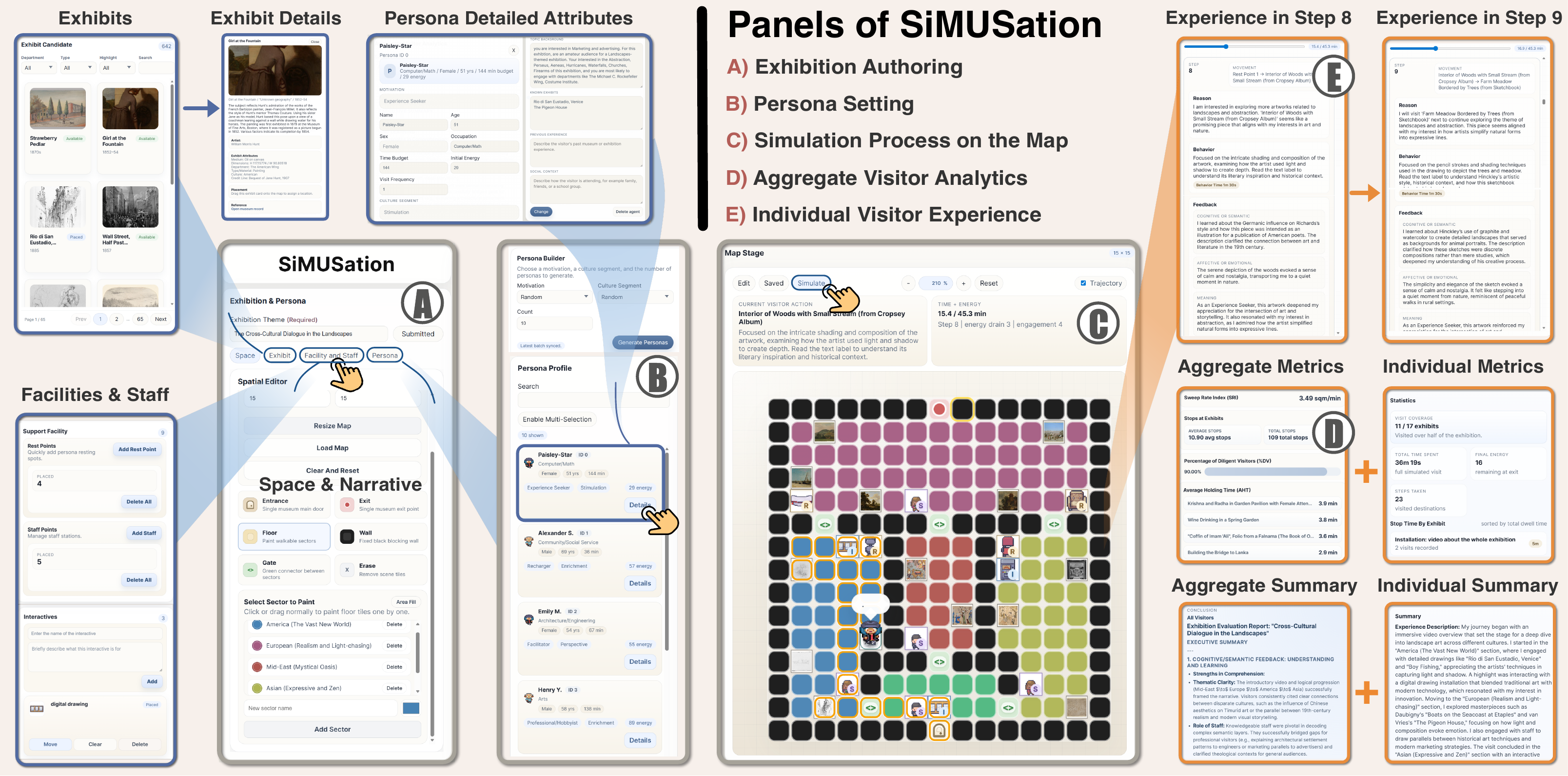}
\caption{\label{fig:interface}
The prototype interface supports an authoring-simulating-iterating workflow through five panels: A) Exhibition Authoring Panel for editing exhibition space, exhibits, facilities, and staff; B) Persona Setting Panel for building and editing visitor personas; C) Simulation Process on the Exhibition Map for switching between editing and simulation modes and visualizing visitor trajectories and interactions; D) Aggregate Visitor Analytics Panel for group-level summaries and behavioral metrics; and E) Individual Visitor Experience Panel for step-by-step inspection of each visitor's behaviors and responses.}
\Description{This figure shows the prototype interface of SiMUSation, which supports an authoring-simulating-iterating workflow through five coordinated panels. Panel A, the Exhibition Authoring Panel, is used to edit exhibition space, exhibits, facilities, and staff. Panel B, the Persona Setting Panel, supports the creation and editing of visitor personas. Panel C visualizes the simulation process on the exhibition map, allowing users to switch between editing and simulation modes and inspect visitor trajectories and interactions. Panel D, the Aggregate Visitor Analytics Panel, presents group-level summaries and behavioral metrics. Panel E, the Individual Visitor Experience Panel, enables step-by-step inspection of each visitor's behaviors and responses.}
\vspace{-5mm}
\end{figure*}
\section{SiMUSation}
Drawing on the derived design goals, we proposed SiMUSation~(\autoref{fig:teaser}), a framework with persona-driven visitor experience simulation to support early-stage museum exhibition design, and developed a prototype based on it. 
\revise{The framework comprises four main components: visitor persona agent construction (DG1, DG2), exhibition spatial and narrative authoring (DG2, DG3), visitor experience simulation and representation (DG2), and multi-granular feedback for insight-driven iteration (DG2, DG3). These components support persona building, dynamic simulation, interpretable feedback, and interactive design refinement, helping designers understand visitor experience and revise exhibition plans.}
In this section, we detail SiMUSation’s framework~(\autoref{fig:teaser}) and prototype~(\autoref{fig:interface}).

\subsection{Visitor Persona Agents Construction}\label{frame:persona}
\subsubsection{Structuring Personas via the Contextual Model:}\label{frame:persona:model}
\revise{To support persona construction that reflects real-world visitor diversity (DG1)}, we ground SiMUSation in Falk and Dierking’s Contextual Model of Learning~\cite{falk2016museum}, which frames museum experience through the interaction of Personal, Sociocultural, and Physical contexts across the phases before, during, and after a visit (DG2). In this component, we use the model to structure visitor attributes that can be specified prior to the visit. Accordingly, the persona agent captures pre-visit characteristics from the Personal, Sociocultural, and Physical dimensions on the visitor side~(\autoref{fig:teaser}.1.a). Contextual factors that emerge during the visit, such as the exhibition environment and its interaction with visitors, are incorporated later in the experience simulation pipeline described in~\autoref{frame:experience}. Specifically, SiMUSation structures persona agents through three dimensions:

\noindent\textbf{1. The Personal Dimension:}
This dimension captures individual characteristics that shape the unique needs and assets a visitor brings to a visit, as well as the self-fulfillment they seek, which influences their visiting experience. The specific attributes include:
\vspace{-1mm}\begin{itemize}[label={}, leftmargin=*]
\item \person{\texttt{\bfseries Identity-related motivation}}: 
\revise{To reflect the diverse primary reasons for visiting in a structured way, we incorporate Falk’s framework~\cite{falk2016identity}, which argues identity-related motivation largely shapes visitor experience and categorizes visitors as \textit{Explorers, Facilitators, Professionals/Hobbyists, Experience Seekers,} or \textit{Rechargers}.}
\item \person{\texttt{\bfseries Prior knowledge}}: The visitor’s familiarity with the exhibition or related domain expertise before the visit (e.g., related exhibits, history, background).
\item \person{\texttt{\bfseries Prior experience}}: Relevant previous museum-going experiences that may shape expectations and comparison.
\item \person{\texttt{\bfseries Interests}}: Specific themes, topics, or types of exhibits that the visitor is particularly drawn to.
\end{itemize}
\vspace{-1mm}

\noindent\textbf{2. The Sociocultural Dimension:}
This dimension captures the social and cultural background through which a visitor engages with an exhibition. It includes both a broader cultural background and an immediate social setting. The specific attributes include:
\vspace{-1mm}\begin{itemize}[label={}, leftmargin=*]
\item \socio{\texttt{\bfseries Cultural segment}}: 
\revise{An audience segmentation system widely used in culture and heritage organizations, characterized by cultural background and social identity, and situating individual perspectives within a broader social and cultural context~\cite{mhm_culture_segments}.}
\item \socio{\texttt{\bfseries Relationship}}: whether the visitor attends alone or with others, such as friends or family.
\end{itemize}
\vspace{-1mm}

\noindent\textbf{3. The Physical Dimension:}
Within persona construction, this dimension focuses on visitor-side physical conditions that can be specified before the visit, rather than on exhibition-side physical factors. These attributes capture physical factors that may impact the experience. The specific attributes include:
\vspace{-1mm}\begin{itemize}[label={}, leftmargin=*]
\item \phys{\texttt{\bfseries Expected visiting duration}}: The approximate amount of time the visitor plans or is willing to spend in the exhibition.
\item \phys{\texttt{\bfseries Stamina}}: The visitor’s general physical endurance for moving through and attending to an exhibition over time~\cite{gilman1916museum}.
\end{itemize}
\vspace{-1mm}

These attributes provide a structured basis for representing visitor diversity in SiMUSation. However, directly exposing all of them as open-ended configuration variables would make persona construction difficult to manage in practice. To retain descriptive richness while supporting interpretability and scalability, we ground persona construction in audience survey data, as described below.

\vspace{-2mm}\subsubsection{Grounding Persona from Audience Surveys}\label{frame:persona:data}
While the attribute set outlined above enables comprehensive visitor characterization, exposing users to too many variables can increase cognitive burden and affect the scalability and reliability of persona construction.
To address this challenge, prior market research suggests that a two-dimensional ($2 \times 2$) framework can offer a structured yet flexible way to understand target audiences~\cite{pollock2012give,lowy2010power}. 
Building on this rationale, SiMUSation adopts an approach centered on two anchor dimensions: \person{\texttt{\bfseries Identity-related motivation}} and \socio{\texttt{\bfseries Cultural segment}}. 
\revise{We selected these two attributes because formative study participants highlighted them as particularly important for shaping the visitor experience, while prior work suggests that identity-related motivation largely influences the museum experience~\cite{falk2016identity}, and cultural segments offer a practical representation widely adopted in cultural and heritage contexts~\cite{mhm_culture_segments}.} 
Thus, these two attributes inform visitors’ exhibition expectations through both personal orientation and sociocultural positioning.

On this basis, the designer can first specify the two anchor dimensions and the number of personas to define the target audience. The remaining attributes are then populated by sampling empirical distributions derived from culture- and museum-related audience surveys (see supplemental materials). Using survey-based distributions instead of LLM generation reduces hallucinated or weakly supported attribute combinations while keeping persona construction practical, interpretable, and grounded, while preserving within-group variation. Designers can also manually refine personas for target audiences or edge cases using the attributes above.

\vspace{-5mm}\subsection{Exhibition Spatial and Narrative Authoring}\label{frame:exhibit}
Early-stage exhibition design is an inherently iterative process that requires flexible, experience-driven adjustments. 
\revise{To support this process (DG3), SiMUSation provides an interactive authoring environment that enables designers to rapidly create and revise exhibition plans. The system also translates these plans into a computational representation that LLM-powered agents can perceive and reason about, enabling dynamic simulation of visitor experience within this environment (DG2).}

\revise{Based on F3, we abstracted the authoring canvas into a 2D grid-based canvas with topology, reducing the LLM's reasoning overhead while preserving the critical exhibition contexts that influence the visitor experience: routing paths, spatial connectivity, and encounters with narratives and exhibition content.}
The authoring process is organized into three functional layers that provide the simulation with exhibition-related context~(\autoref{fig:teaser}.1.b). The first layer defines the space, including walls, entrances, exits, and passage points, as well as floor areas that indicate functional and narrative sectors. All these elements establish the exhibition's accessible boundaries and provide the spatial and semantic context for visitor movement. The second layer consists of exhibits placed within these defined zones. Each exhibit is represented through multi-modal information, including images, content and background descriptions, and other attributes such as artist and medium. The third layer includes facilities and staff, such as interactive installations, resting areas, and staffs, which represent environmental and operational factors that may influence visitor experience.
After authoring, SiMUSation converts the configured layout into a JSON representation, translating the exhibition plan into a visiting simulation.

\vspace{-2.5mm}\subsection{Visitor Experience Simulation and Representation}\label{frame:experience}
\begin{figure}[!t]
  \centering
  \vspace{-3mm}
  \includegraphics[width=0.96\linewidth]{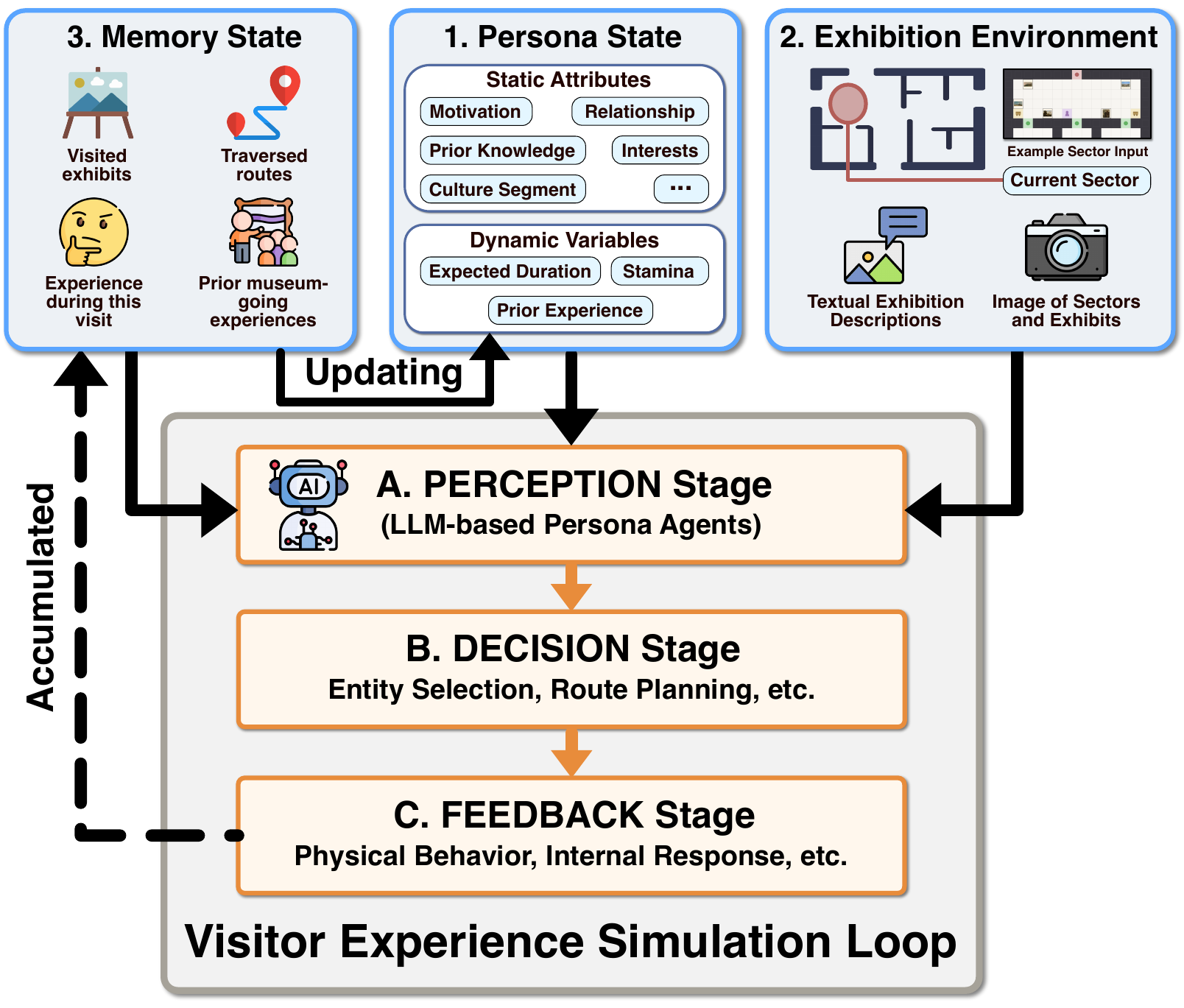}
  \caption{Perception-Decision-Feedback simulation loop in SiMUSation, where each LLM-driven visitor agent updates its behavior and responses over time based on persona state, situated environmental input, and memory state.}
  \Description{This figure illustrates the Perception-Decision-Feedback simulation loop in SiMUSation. In this loop, each LLM-driven visitor agent continuously updates its behaviors and responses over time based on three factors: the current persona state, situated input from the environment, and dynamic memory state. The figure emphasizes how visitor experience is modeled as a dynamic and stateful process rather than a one-time response.}
  \label{fig:pipeline}
    \vspace{-5mm}
\end{figure}

Building on the persona structure introduced above, SiMUSation models visitor experience as an evolving process shaped by the interaction of personal, sociocultural, and physical contexts throughout the visit. Rather than treating personas as static profiles, the framework uses these factors as simulation inputs to account for how LLM-driven visitors attend to exhibits, move through the exhibition space, and respond to what they encounter. To achieve this, we design the experience simulation as a multi-modal Perception-Decision-Feedback loop~(\autoref{fig:pipeline}) and translate the simulation process into a dual-layer representation,~\autoref{fig:teaser}.2 (DG2).

At each step of the simulation loop, the LLM-based persona agent is conditioned on three core input streams. 

\noindent\textbf{1. Visitor persona state}~(\autoref{fig:pipeline}.1) provides the basis for decision making. These include the persona attributes defined in~\autoref{frame:persona} (i.e., 
\person{\texttt{\bfseries Identity-related motivation}}, \person{\texttt{\bfseries Prior knowledge}}, \person{\texttt{\bfseries Interests}}, \socio{\texttt{\bfseries Culture segment}}, \socio{\texttt{\bfseries Relationship}}), which affect expectations, exhibit selection, and route planning. They also include dynamic state variables such as \phys{\texttt{\bfseries Stamina}}~\cite{gilman1916museum} and \phys{\texttt{\bfseries Expected visiting duration}}, which may influence how long visitors continue exploring, how selectively they browse, and whether they seek rest. In addition, \person{\texttt{\bfseries Prior experience}} can be derived from prior simulated visiting experience to model returning visitors. 

\noindent\textbf{2. Situated environmental input}~(\autoref{fig:pipeline}.2) provides the locally available context for perception. SiMUSation limits each visitor's sight to the current sector, including nearby exhibits, facilities, and staff, rather than assuming access to the full exhibition at once. This input is provided through both a layout image of the current sector and structured textual descriptions of accessible entities. When the visitor chooses to inspect a specific exhibit, the agent further receives exhibit-level multi-modal input, including the exhibit image and its descriptive metadata. 

\noindent\textbf{3. Memory state}~(\autoref{fig:pipeline}.3) maintains information accumulated across the visit, such as visited exhibits, traversed routes, and experience during this visit, while also incorporating relevant \person{\texttt{\bfseries Prior experience}}. It supports continuity across steps and enables more coherent sequential behavior and interpretation over time.

Given these inputs, the simulation proceeds iteratively through perception, decision, and feedback. In the perception phase, the visitor receives multi-modal information from the local environment, along with its persona-related state and memory. In the decision phase, the model determines the next action based on these combined inputs, such as moving toward another exhibit, interacting with an installation, asking staff for directions or interpretation, or resting in a designated area. 
In the feedback phase, the consequences of that action update the visitor's state, memory, and subsequent access to information. For example, resting may reduce fatigue, while interaction with exhibits, facilities, or staff may alter later movement, attention, and response. These updated states then form part of the input to the next simulation step.

To make the resulting simulation trace useful for exhibition design, SiMUSation represents visitor experience feedback through two aligned layers: observable physical behavior and internal response~\cite{levasseur1983visiting, gem2026glo, falk2016museum, ILI2025measuring, weng2023agent, zhou2024sotopia}. \revise{This representation draws on prior work and our formative study findings, with further details in the supplemental material.}
\textbf{The first layer} captures visible behaviors that designers could reasonably infer or observe in real settings~\cite{levasseur1983visiting}, including movement and navigation behaviors, information-access and attention behaviors such as viewing an exhibit, interaction and engagement behaviors such as taking notes or sketching, and social behaviors such as speaking with companions. 
All these behaviors make the simulated visit legible as a trajectory through the exhibition space;
\textbf{The second layer} captures the visitor's moment-to-moment internal response during the visit. 
Accordingly, SiMUSation represents internal response across three closely related dimensions: cognitive and comprehension response (e.g., whether the visitor understands the content or can connect it to prior knowledge), affective response (e.g., feelings of interest, confusion, surprise, boredom, or emotional resonance), and meaning- and value-related response~\cite{falk2016museum, ILI2025measuring}.
The latter includes ways in which the visit may contribute to personal, intellectual, social, or physical well-being, such as a sense of identity, reflection, or restoration.

Both layers allow SiMUSation to represent visitor experience not only as a sequence of actions, but also as an interpretable process of situated reasoning and response. For each simulated step, SiMUSation can therefore output both what the visitor does and how the visit is internally experienced, helping designers examine how specific spatial, curatorial, and interpretive choices may influence different audiences throughout the visit.

\vspace{-3mm}\subsection{Multi-Granular Feedback for \revise{Insight-Driven} Iteration}
To support \revise{insight-driven} iteration, SiMUSation follows an overview-to-detail analysis logic~\cite{shneiderman2003overviewdetail} and provides multi-granular quantitative and qualitative feedback from individual visitors to aggregated groups (DG2). This structure allows designers to move between broad patterns and specific cases, while comparing how visitor experience varies across different audiences in the exhibition, thus further supporting iterative refinement and revision (DG3).

From each simulated visit, SiMUSation first extracts individual-level behavioral traces, including visited exhibits and sectors, dwell time, stops, and routes. Based on these traces, it computes behavioral evaluation metrics~\cite{wieland2026evaluate}, such as Average Holding Time and Walkthrough Rate~(\autoref{tab:evaluation_metrics_formula}), to assess attention, circulation, and engagement patterns across personas and visitor groups.
In parallel, SiMUSation summarizes each visitor's qualitative feedback from the full visit trace, including overall impressions, suggestions, and satisfaction~(as detailed in the supplemental material).
These persona-grounded summaries can then be aggregated further for selected visitor groups or the full simulated population, enabling comparisons of shared and divergent responses across audiences.

\subsection{Prototype}
Based on the SiMUSation framework, we developed a working prototype of an LLM-powered interactive system that simulates various visitor personas in a virtual environment as a reference for early-stage exhibition design~(\autoref{fig:interface}).

\subsubsection{Implementation Details}
The SiMUSation prototype is developed as a web-based application. The front-end interface is built using Vite and Vue.js to support responsive and interactive visualizations. The back-end relies on FastAPI to manage system requests and LangChain to orchestrate the iterative perception-decision-feedback simulation loops for the LLM-based agents. We utilize Qwen3.5-Plus~\cite{qwen35blog} as our core LLM for its vision-language understanding and spatial reasoning capabilities, enabling it to process multi-modal inputs such as textual descriptions and exhibit images.

The simulation loop is mainly driven by three LLM-based agents. The \textbf{Visitor Agent} governs each simulated persona through two chained sub-stages and one summary stage: a \textit{Navigation Decision} step that decides which entity in environment to approach next based on the visitor's profile, current position, and prior experience of before and during visiting, and an \textit{Experience Feedback} step that generates the visitor's concrete behavior, responses, and inner attribution upon arriving at the chosen exhibit. When ending of the visiting, the agent will \textit{Summary} the experience. The \textbf{Staff Agent} simulates docent or guide interactions when a visitor persona initiates a question, producing context-aware dialogue grounded in the exhibit's curatorial narrative.
Finally, the \textbf{Analyst Agent} synthesizes each visitor's full trajectory and feedback into a structured post-visit summary that highlights key experiential patterns and potential design issues. All agents receive structured output formats to ensure machine-parseable responses that can be directly rendered by the front-end. Further implementation details and prompts are provided in the supplemental material.

\begin{figure*}[htbp!]
\centering
\includegraphics[width=\textwidth]{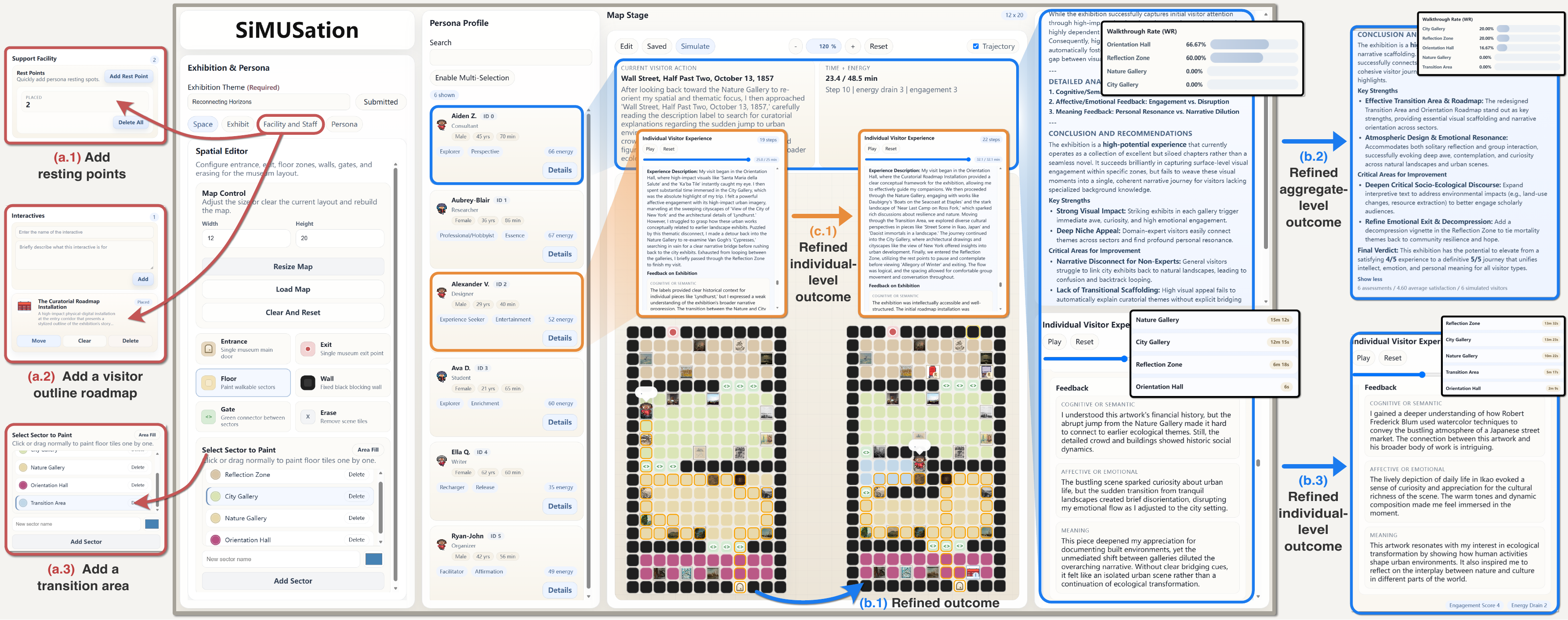}
\caption{\label{fig:userscenario} An early-stage exhibition design scenario in SiMUSation, illustrating iterative reasoning through persona construction, simulation feedback analysis, and design refinement: (a) refinement actions; (b) and (c) refined results. }
\Description{This figure shows an early-stage exhibition design scenario in SiMUSation. The figure presents iterative design reasoning through persona construction, simulation feedback analysis, and design refinement. Panel (a) shows the refinement actions applied to the exhibition layout. Panels (b) and (c) show post-refinement simulation results, indicating improved narrative flow, smoother transitions between sectors, and stronger engagement across visitor groups.}
\vspace{-3mm}
\end{figure*}

\subsubsection{Interface}
The prototype interface consists mainly of five panels~(\autoref{fig:interface}): Exhibition Authoring Panel; Persona Setting Panel; Simulation Process on the Exhibition Map; Aggregate Visitor Analytics Panel; Individual Visitor Experience Panel to support an iterative authoring-simulating-iterating workflow, enabling designers to easily test and refine their design plans:

\noindent\textbf{Exhibition Authoring Panel (DG3)}: 
This panel supports authoring for the exhibition. Users first specify the exhibition theme, and then switch among three authoring views corresponding to the layers in~\autoref{frame:exhibit}: \textit{space view}, \textit{exhibit view}, and \textit{facilities and staff view}~(\autoref{fig:interface}.A). They are organized around a shared Map Panel, allowing users to configure the exhibition context for simulation.

In the \textit{space view}, users can upload a floor plan or edit the map directly by placing entrances, exits, floors, gates/passages, and walls. Designers can name each sector for narrative or functional roles and assign floor colors to distinguish areas. In the \textit{exhibit view}, users can browse exhibits by filtering with department, type, and highlight tags, or by search. Clicking an exhibit card reveals its detailed information, and selected exhibits can be placed onto the Map Panel. In the \textit{facilities and staff view}, users can place rest points and staff locations, and configure the content of interactive facilities before positioning them on the Map Panel. These views support the structured authoring of the spatial, semantic, and operational context required for visitor simulation.

\noindent\textbf{Persona Setting Panel (DG1, DG2)}: 
This panel includes a \textit{persona building view} and a \textit{persona profiles view}~(\autoref{fig:interface}.B). In the persona building view, users can generate a specified number of personas based on selections from the two anchors in~\autoref{frame:persona:data}, or generate them fully at random. The persona profiles view displays each persona's basic information. Clicking a persona card reveals the detailed attributes described in~\autoref{frame:persona:model}, which users can further edit. Users can also delete personas in batches.

\noindent\textbf{Simulation Process on the Exhibition Map (DG2, DG3)}: 
This panel supports both design-plan editing and the visualization of the visiting simulation process on the exhibition map~(\autoref{fig:interface}.C). Users can switch between \textit{edit} and \textit{simulate} modes through the controls at the top of the panel. In \textit{simulate} mode, the panel visualizes each visitor's movement trajectory and interactions with entities in the environment. It also displays the current simulation step with the corresponding behavior, time, and stamina state.

\noindent\textbf{Aggregate Visitor Analytics Panel (DG2, DG3)}: 
Following an overview-to-detail analysis mantra, this panel provides aggregated summaries and behavioral metrics for selected visitor groups or the full simulated population~(\autoref{fig:interface}.D). Users can select groups through a drop-down menu based on the two persona anchors. The panel presents group-level summaries of visitor satisfaction, overall feedback, perceived strengths, and suggestions, with evaluation metrics including Visitor Count (VC), Average Holding Time (AHT), Attracting Power (AP), Sweep Rate Index (SRI), Percentage of Diligent Visitors (\%DV), Walkthrough Rate (WR), Stops at Exhibits, and Dwell Time (DT), in~\autoref{tab:evaluation_metrics_formula}. These outputs help users examine overall exhibition performance across different audiences.

\noindent\textbf{Individual Visitor Experience Panel (DG2, DG3)}: 
In the individual panel, this panel presents each visitor's step-by-step experience during the visit with a final summary of the overall experience~(\autoref{fig:interface}.E). It also reports individual behavioral measures, such as visited sectors and time spent, stopped exhibits and dwell time, and total visit duration. Users can pause the simulation shown in the Map Panel or drag the timeline to inspect what the visitor was doing at a specific step. These two panels support both cross-group analysis and case-level inspection of the simulated visitor experience, enabling multi-granular analysis to inform iteration.

\vspace{-2.5mm}\subsection{User Scenario}
\revise{To illustrate how SiMUSation supports design reasoning, we describe an early-stage exhibition scenario. Designer Lily is developing a temporary exhibition on how urban development reshapes human relationships with the natural environment. The goal is to connect scenic imagery, ecological observation, and urban experience through a coherent visitor journey. Lily has a preliminary concept and candidate exhibits, but is unsure whether the arrangement can convey the intended narrative to different visitors.}

\revise{Lily divides the exhibition into four sectors: an \textit{Orientation Hall}, a \textit{Nature Gallery}, a \textit{City Gallery}, and a \textit{Reflection Zone} linking the two perspectives. Exhibits and pathways are then arranged on the map panel to guide visitors from natural observation to urban development and personal reflection~(\autoref{fig:userscenario}).}
\revise{Lily then defines a small set of personas to represent likely audience diversity, such as an \textit{Explorer} interested in ecology and local geography, and an \textit{Experience Seeker} visiting casually for visual enjoyment. Using SiMUSation’s persona construction, Lily reviews the generated profiles and makes minor adjustments to match the expected audience.}

\revise{After running the simulation, Lily first examines aggregate feedback~(\autoref{fig:userscenario}.b.2). The results show that exhibits in the \textit{Orientation Hall} lack clear guidance, while the \textit{Reflection Zone} has a high walk-through rate and few sustained stops. And the summaries suggest that visitors already interested in environmental or city topics still perceive the sectors as isolated, whereas casual visitors often experience the exhibition as visually appealing but loosely connected scenes, resulting in limited attention to the \textit{Reflection Zone}. The design thus partially captures some audiences' attention, but is less effective in sustaining narrative continuity and meaning-making.}

\revise{To understand why, Lily inspects individual simulated visits. \textit{Alexander} (ID 2) spends substantial time in the \textit{City Gallery} and shows strong affective engagement with urban imagery, but expresses confusion and fatigue about how these works relate to earlier landscape exhibits~(\autoref{fig:userscenario}.c.1). \textit{Aiden} (ID 0) follows the intended route more closely, yet feels the shift from the \textit{Nature Gallery} to the \textit{City Gallery} is abrupt~(\autoref{fig:userscenario}.b.3).}

\revise{Based on these findings, Lily revises the plan by adding an outline map into the \textit{Orientation Hall}, adding a \textit{Transition Area} between the \textit{Nature Gallery} and \textit{City Gallery}, and placing a resting point near the entrance to the \textit{Reflection Zone}~(\autoref{fig:userscenario}.a). Based on the refinement, the feedback of the second simulation shows these changes improve narrative flow continuity and sustain engagement across these audience groups~(\autoref{fig:userscenario}.b, \ref{fig:userscenario}.c) .}
\section{Evaluation}
To evaluate the reliability and usability of SiMUSation, we conducted a user study to explore the following two research questions:
\begin{itemize}
    \item[\textbf{RQ1}:] How do designers perceive the use of LLM-driven visitor personas as proxies for real audiences?
    \item[\textbf{RQ2}:] How does the iterative simulation-feedback framework support the early-stage exhibition design?
\end{itemize}

\subsection{Methods}
\noindent\textbf{Participants.}
We recruited 12 participants aged 19-33 years ($M=23.00$; 6 females and 6 males; see details in~\autoref{table:user participants}) through campus forums and referral-based outreach. All participants had at least one year of exhibition design experience, including nine students in related fields and three practitioners. The user study consisted of a 2-hour session with our prototype, conducted either in person or remotely via Zoom, and participants received 100 CNY (approximately USD 14) as compensation. To evaluate the effectiveness of users' refinements with SiMUSation, we also invited four domain experts with backgrounds in design and fine arts and rich curation experience to review the users' results of each study. 

\noindent\textbf{Task and Data.}
We asked participants to design an exhibition around the theme \textit{cross-cultural dialogue in landscapes}. To support this task, we curated a set of open-access information on 600+ artworks related to landscape representation across different cultures from The Metropolitan Museum of Art collection\footnote{\url{https://www.metmuseum.org/art/collection}}. We obtained the artworks' metadata and images through the museum's official API\footnote{\url{https://metmuseum.github.io/}} and converted them into the format required by SiMUSation. Given a predefined exhibition space, participants were asked to select around 20 exhibits, arrange them in the layout, and construct a coherent narrative for the exhibition.

\noindent\textbf{Procedure.}
Participants were asked to familiarize themselves with the exhibition theme and dataset one week before the study. On the study day, each session began with a 15-minute introduction to the study background and relevant domain knowledge, followed by a 30-minute system walkthrough and free exploration to familiarize with the functions and interactions. Participants then completed a 50-minute design task, progressing from initial exhibition authoring to simulation, iteration, and refinement. Afterward, they spent 10 minutes completing questionnaires and 15 minutes in a semi-structured interview. 
\revise{We logged each participant’s initial designs (v1, authored without simulation) and final designs (v2, revised with simulation), and invited four experts to review and score the results.}
For each participant, the experts evaluated the two designs in pairs. To support blind review and reduce bias, we shuffled both the order of the design pairs across participants and the placement order of the initial and final designs within each pair. These mappings were known only to the research team during scoring.

\noindent\textbf{Measurements.}
To assess users' perceptions of SiMUSation, we evaluated its usability using the System Usability Scale (SUS)~\cite{brooke1996sus}. We also measured perceived reliability and early-stage design support using 7-point Likert-scale items. The measurements were as follows. For reliability and acceptance of persona simulation: 
\vspace{-0.5mm}\begin{itemize}[leftmargin=*]
\item  \textbf{Persona differentiation}: whether the framework captured differences in exhibition experience across different personas; 
\item  \textbf{Behavioral plausibility}: whether the simulated behavior aligned with expected visitor behavior in real exhibition settings; 
\item  \textbf{Response realism}: whether simulated internal responses, such as emotions and narrative understanding, appeared natural and consistent with visitor personas; 
\item  \textbf{Experience explainability}: whether the simulation helped participants understand the reasons behind visitors' feedback. 
\end{itemize}
\vspace{-0.5mm}
For the effectiveness of the SiMUSation, we assessed: 
\vspace{-0.5mm}\begin{itemize}[leftmargin=*]
\item \textbf{Design insight}: whether the simulation revealed design blind spots or potential problems overlooked in initial planning; 
\item  \textbf{Decision support}: whether outputs such as trajectories and individual feedback provided useful support for revising layout or narrative structure; 
\item \textbf{Design confidence}: whether using SiMUSation increased confidence in the design compared with relying only on intuition; 
\item  \textbf{Iteration effort}: whether SiMUSation enabled easy and low-cost testing of different layouts and exhibit combinations; 
\item  \textbf{Workflow optimization}: whether integrating editing and experience simulation improved early-stage exhibition planning.
\end{itemize}
\vspace{-0.5mm}
For expert evaluation, each pair of designs from the same participant was reviewed separately using 7-point Likert-scale ratings on the following criteria: 1) \textbf{Narrative coherence}: whether the cross-cultural theme was presented through a clear and reasonable narrative logic; 2) \textbf{Spatial organization}: whether spatial exhibit order and circulation were natural and well organized; 3) \textbf{Facility placement suitability}: whether the locations of facilities and staff were suitable in the exhibition layout.

\begin{figure}[!htbp]
\vspace{-3mm}
\centering
\includegraphics[width=0.4\textwidth]{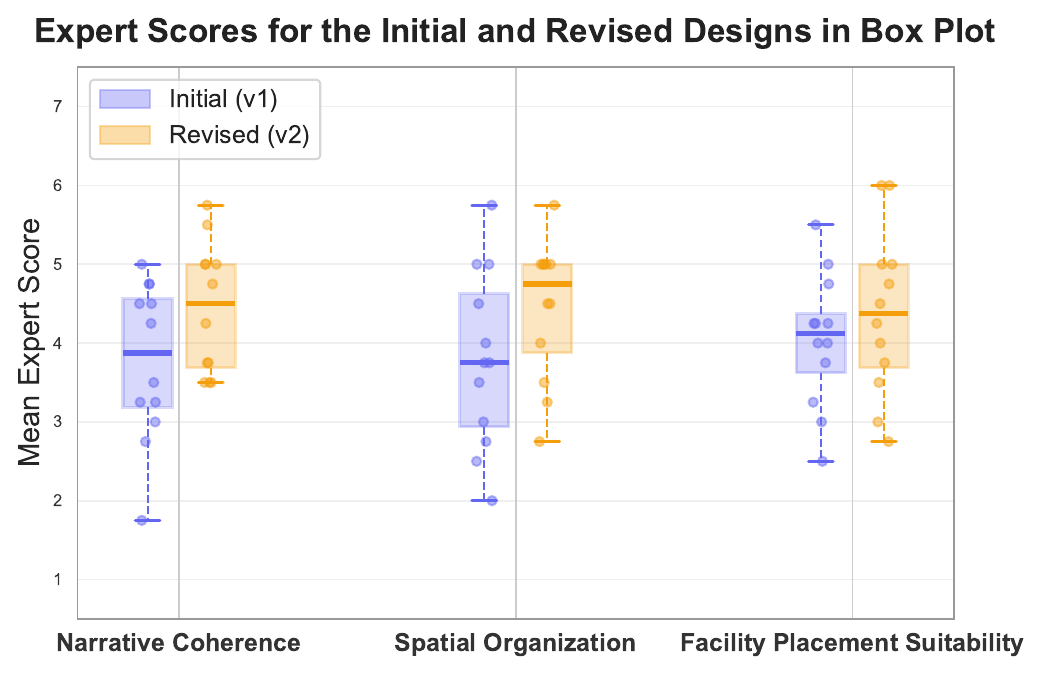}
\caption{\label{figure:expert score}
Four experts' ratings of the design from twelve participants between the initial and revised designs.}
\Description{This figure shows four experts' ratings of the designs produced by twelve participants, comparing the initial designs with the revised designs.}
\vspace{-4mm}
\end{figure}

\subsection{Results}
Based on questionnaire responses, post-study interviews, and expert evaluations of participants' design outcomes, the results indicate that SiMUSation was positively received in terms of both usability and perceived usefulness for early-stage exhibition design. The SUS analysis yielded a mean score of 79.38 (SD = 7.78), suggesting a good level of usability. Beyond overall usability, the findings further suggest that participants generally found the LLM-driven personas plausible and their responses interpretable. Participants also considered integrating simulation with feedback into the workflow as helpful for identifying design issues and supporting refinement. 
These perceptions were further reflected in the expert assessments, in which revised designs generally received higher scores than the initial versions on narrative coherence, spatial organization, and facility placement suitability~(\autoref{figure:expert score}).
\revise{The inter-rater agreement among experts was ICC(2,1) = 0.83 for v1 and ICC(2,1) = 0.77 for v2.}

\begin{figure}[!htbp]
\vspace{-4mm}
\centering
\includegraphics[width=0.4\textwidth]{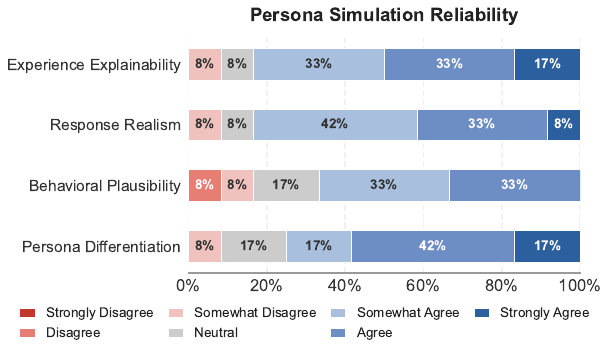}
\caption{\label{figure:persona scale}
Twelve participants' ratings of the questionnaire related to the reliability of SiMUSation.}
\Description{This figure shows the twelve participants' ratings on the questionnaire assessing the reliability of SiMUSation.}
\vspace{-4mm}
\end{figure}

\subsubsection{Plausibility and Explainability of LLM-Driven Personas}\label{eval: result persona}
Participants generally regarded the simulated personas as credible proxies for exhibition audiences, and the quantitative results showed consistently positive evaluations across all four measures~(\autoref{figure:persona scale}). For \textbf{persona differentiation} (M=5.42, SD=1.19), participants perceived the generated visitors as meaningfully distinct rather than interchangeable. This perception was not only based on demographic variation, but also on exhibition-related dimensions such as prior experience and interests. For example, U1 noted that motivations and cultural segments were core dimensions, allowing them to generate personas in batches and then further customize extreme or boundary cases one by one. In this way, audience consideration did not rely solely on personal intuition.

For \textbf{behavioral plausibility} (M=4.75, SD=1.23), the overall evaluation remained positive. Participants generally considered behaviors sufficiently reasonable for judging visitor flow and attention distribution in early-stage planning. U12 commented that paths and dwell time were relatively objective in exhibition settings and therefore aligned well with real design concerns. At the same time, some participants pointed out limits to realism. For instance, U4 noted that other behaviors, such as interaction behaviors, need greater richness from real visitors. Notably, these concerns were typically framed as limitations in modeling complex human behavior rather than as a rejection of the approach itself. As U8 suggested, ``\textit{even if some aspects of real-world experience are somehow simplified, the current framework remains useful for early-stage design}.''

Participants gave more positive evaluations to \textbf{response realism} (M=5.25, SD=1.01), and especially to \textbf{experience explainability} (M=5.42, SD=1.11). The findings suggest that participants valued SiMUSation because it showed not only what a persona did, but also why each decision was made. Several participants emphasized that the combination of trajectory visualization, internal responses, and step-by-step reasoning made the simulated experience easier to understand and, in turn, more credible. As U9 said, ``\textit{I found a repeated viewing pattern initially seemed confusing, but became reasonable after considering the persona's profile as an energy explorer and stepwise feedback.}'' 
This indicates that explainability directly influenced whether participants regarded a persona as trustworthy.
In this respect, the internal responses were particularly important. Participants appreciated SiMUSation's ability to display immediate reactions across multiple dimensions and relate them to persona-specific states. U11, for instance, noted that they read the internal responses most frequently and found them the most useful, particularly for assessing whether the persona's thoughts aligned with the ideas they intended to convey. Rather than producing generic comments on the exhibition, these responses were often interpreted as grounded in each persona's interests and background.
\revise{Despite these strengths, participants also noted risks related to LLM sycophancy and hallucination. U7 and U4 found the aggregate-level summary overly positive and insufficiently critical, even when some layouts were intentionally confusing for visitors. U2 also questioned whether real visitors' motivations would ever be so explicit.}

Overall, participants' trust in the personas depended less on whether every action fully reproduced real human behavior, and more on whether the simulated experience was interpretable, insightful, consistent with the persona, and supported by a traceable causal chain. This perceived credibility was also important for later use of SiMUSation, as participants were willing to treat simulated visitor experiences as actionable evidence precisely because they could inspect both movement trajectories and internal responses.

\begin{figure}[!htbp]
\vspace{-4mm}
\centering
\includegraphics[width=0.4\textwidth]{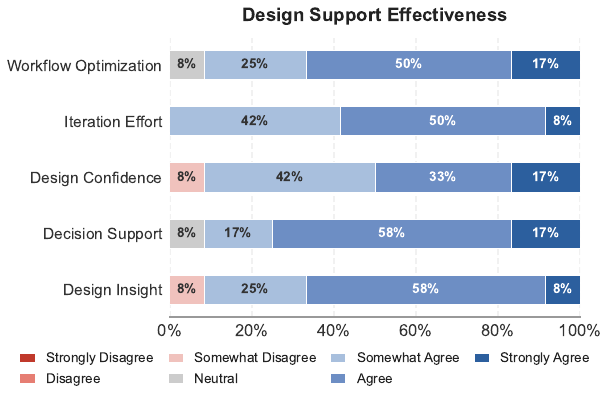}
\caption{\label{figure:design scale}
Twelve participants' ratings of the questionnaire related to the effectiveness of SiMUSation.}
\Description{This figure shows the twelve participants' ratings on the questionnaire assessing the effectiveness of SiMUSation.}
\vspace{-5mm}
\end{figure}

\subsubsection{Support for Early-Stage Exhibition Design through Iterative Simulation and Feedback}\label{eval: result design}
The results also suggest that participants found SiMUSation useful for early-stage exhibition design, especially in revealing overlooked issues, supporting revision, reducing iteration effort, and increasing confidence in decision-making~(\autoref{figure:design scale}). This indicates that SiMUSation is not only a simulation tool, but also an iterative design aid for early-stage design.

For \textbf{design insight} (M=5.58, SD=0.95), participants reported that the simulation helped expose design blind spots that were not apparent in their initial plans. U3 noted from the visitor trajectories that, although the cultural sections had been arranged according to their real-world geographic positions, many personas mainly attended to the central and rear spaces while overlooking the side areas, suggesting that this spatial logic was not immediately clear and that stronger guidance was needed. Similarly, U4 reflected after reading the feedback that the overview area, which they had initially designed around a highlighted exhibit to serve as a strong hook, was not effective enough in helping visitors both grasp the overall exhibition structure and transition into the main theme.

For \textbf{decision support} (M=5.83, SD=0.80), SiMUSation supports design revision by combining aggregated and individual-level evidence across both quantitative and qualitative forms. At the macro level, U9 noted that visualized metrics made overall trends and attention hotspots easy to identify. At the micro level, U12 emphasized that individual behaviors and thoughts helped them better take the visitor's perspective to review the design. This combination, in which high-level patterns indicated where problems occurred, and persona-level evidence helped explain why, also strengthened participants' overall \textbf{design confidence} (M=5.50, SD=1.04). U7 suggested that because the simulated persona behavior was rational and explainable, it increased their confidence for designs rather than personal intuition. This finding is also consistent with the role of explainability discussed in~\autoref{eval: result persona} in supporting understanding and decision-making.

The results for \textbf{iteration effort} (M=5.67, SD=0.62) and \textbf{workflow optimization} (M=5.75, SD=0.83) further indicate clear advantages. Compared with traditional approaches such as writing outlines, sketching, or manually building 3D models, SiMUSation supported a tightly integrated and relatively low-cost edit-simulate-analyze-revise loop. U11 noted that, because layout editing and simulation feedback were integrated into the same environment, designers could avoid labor-intensive manual drafting. This aligns with U2's observation that the tool made both layout modification and simulation results visible and editable. 

Expert evaluations of the initial and revised designs provided complement whether support early-stage design. The results showed that revised designs mainly received higher scores than the initial versions across all three evaluation dimensions. Among these, the improvements in \textbf{narrative coherence} ($\Delta$, M=0.67, SD=0.98) and \textbf{spatial organization} ($\Delta$, M=0.65, SD=0.97) were especially notable. These findings are consistent with participants' reports that simulation visualizations and step-by-step feedback helped them identify and address imbalanced circulation, movement bottlenecks, and neglected areas, thereby improving the coherence and integration of the exhibition content and space.

Overall, the main value of SiMUSation in early-stage exhibition design lies in its support for a practical, visitor-centered iterative loop. It helped designers identify design problems, explore alternatives at relatively low cost, and understand diverse visitor experiences for decision making, thereby translating these insights into improvements in spatial organization and narrative coherence.

\section{Discussion}
Through the design and development of SiMUSation, informed by findings from the user study, we derive implications that extend beyond the framework itself for the design of LLM-supported tools in culture- and space-related creative tasks, such as exhibition design and planning. We also reflect on the current limitations of the framework and outline directions for future work.

\vspace{-1.5mm}\subsection{Design Implications}
\subsubsection{From Generator to Experiential Reflection}
Many AI tools for design still operate according to their end-to-end nature, producing relatively complete outputs directly from user prompts. While this approach can improve efficiency, heavy reliance on automated generation may weaken user experience and ownership~\cite{buschek2021UX,draxler2024ownership}, and may also constrain human creativity and critical reflection~\cite{HABIB2024100072, doshi2024novel}. Prior research on human-AI co-creation has examined both the role of AI~\cite{gao2024hai, zhou2024hai} and the workflow across tasks~\cite{zhang2025mixed, shaikh2025mixed}, in which users express intent while AI collects and understands it, then contributes to content creation, followed by user review, decision, and revision.
Our findings suggest that this model may be less suitable for exhibition design, where subjectivity, cultural interpretation, and narrative construction are central. Participants did not primarily expect AI to generate or modify layouts. Instead, they valued AI as a support for experiential reflection, enabling them to examine their ideas from different visitor perspectives and to provide feedback on the design. What mattered was not automated production, but the system's ability to offer diverse, plausible, and reflective responses that supported later refinement.
This also suggests that, for other user-centered creative tasks, AI-supported systems may benefit from shifting from generation-driven to feedback-driven support. In this role, AI acts less as a designer and more as a critical reviewer that supports reflection while preserving user agency.

\vspace{-1mm}\subsubsection{Towards Situated Agents for Design}
Conventional AI-based design assistants are strong in semantic understanding and reasoning but often remain detached from physical constraints~\cite{xu2024physical,liu2025physical}. For design tasks involving space and movement, AI support may benefit from going beyond prompt-response generation.
We suggest considering embodied or situated modeling in agent design, where perception, decision-making, and feedback are continuously integrated with internal states, environmental conditions, and dynamic memory~\cite{park2023simulation,guo2026embodied,fung2025embodied}. In this way, behavior is shaped by both the present context and prior interactions over time.
Such stateful agents provide more realistic feedback for user-centered design, helping designers surface issues in narrative flow, spatial sequence, and experiential continuity that static generation cannot capture.

\vspace{-1mm}\subsubsection{Explainability and Trustworthy AI}
Our findings show that explainability supported transparency and understanding in the use of AI personas. By making visible why a persona behaved as it did, SiMUSation allowed designers to inspect both the observable process and the underlying rationale. This helped participants interpret the simulated experience from the visitor's perspective and relate it to their own design intentions.
\revise{However, while such explainability builds immediate trust~\cite{ferrario2022XAI, chander2025XAI}, fostering grounded, lasting trustworthiness~\cite{atf2026XAI, ferrario2022XAI} requires recognizing that explainability should not be conflated with real-world accuracy. Although SiMUSation provides traceable rationales for reflection, these are model-generated accounts rather than ground-truth evidence of human cognition. To prevent over-reliance on simulated feedback, future AI design aids must move beyond merely explaining behaviors to explicitly communicating AI risks and uncertainties.}

\vspace{-1.5mm}\subsection{Limitations and Future Work}
Our work has several limitations that point to directions for future research. First, SiMUSation currently relies on a 2D representation, which simplifies some properties of real exhibition environments. 
Extending the system to 3D simulation environments may support richer spatial perception (e.g., 3D sightlines, lighting) and more immersive forms of evaluation.
\revise{Second, although persona generation is grounded in empirical data, LLM-powered agents may reflect clichés, sycophancy, or overly positive judgments. This may reduce the system's ability to provide sufficiently critical feedback in edge cases. To mitigate this,  future work could explore fine-tuned models and broader visitor identity settings to improve the rigor and diversity of simulated responses. It could also introduce more robust principles and computational models beyond the current dual-layer representation to better align the simulation with the logic of human perception.}
Third, real exhibitions are not only individual experiences but also social settings. The current framework mainly models individual visiting while incorporating basic social relationships, such as peer influence. It does not yet capture richer dynamic multi-agent interactions, such as crowd formation. Modeling such processes may enhance ecological validity and reveal emergent spatial-use patterns.
Finally, the current pipeline requires re-running the simulation after even minor changes. Incremental simulation methods that update only locally affected agent states to better support iterative author-simulate-iterate workflows.

\section{Conclusion}
In this study, we present SiMUSation, an interactive framework that uses LLM-driven visitor personas to simulate visiting experiences and support early-stage museum exhibition design. SiMUSation combines diverse visitor profiles with authored exhibition contexts to model visitor experiences, represented through physical behaviors and internal responses within the exhibition. Through these simulated personas, designers can inspect how visits unfold, obtain multi-faceted feedback from diverse audience perspectives, and iteratively refine spatial and narrative plans before physical implementation. The user study suggests that the virtual personas can serve as a reference for exhibition audiences and help users identify overlooked design blind spots, thereby enriching and supporting the early-stage exhibition design process. SiMUSation also shows the potential of LLM-powered simulation for audience-informed design and evaluation tasks.

\begin{acks}
We sincerely thank the participants in our formative study and user study for sharing their thoughts and suggestions on SiMUSation, and the four experts for their careful ratings and feedback. We are also grateful to all reviewers for their valuable insights and feedback. We acknowledge the partial use of LLMs to assist in the writing process. The LLMs were employed as a tool for polishing the manuscript to enhance the clarity and quality of the text.
\end{acks}

\clearpage
\bibliographystyle{ACM-Reference-Format}
\bibliography{reference}


\appendix

\section{Formative Study}\label{A:formative}

\subsection{Brief Task}\label{A:brief task}
Participants were asked to complete a brief exhibition design task situated in a realistic museum setting. The scenario was set in the first-floor red zone of the Denon Wing at the Louvre\footnote{\url{https://collections.louvre.fr/}}. The exhibition theme was \textit{Renaissance painting}. To support the task, we provided each participant with the floor plan and 50 related paintings from the Louvre collection.

Participants were asked to arrange the artwork cards on the floor plan as an initial exhibition concept. The task focused on curatorial thinking rather than technical feasibility, encouraging them to use grouping, sequence, and spatial placement to organize the selected works into one or more coherent narratives, ideas, or concepts related to the Renaissance.

\begin{table}[H]
\caption{Details of participants interviewed in the formative study, including their professional backgrounds and museum or exhibition practice experience.}
  \Description{This table demonstrates the summary of participants interviewed in the formative study.}
  \label{table:formative participants}
\resizebox{0.48\textwidth}{!}{
\renewcommand\arraystretch{0.9}
\begin{tabular}{ccc}
\hline
ID   & Background                          & Domain Experience\\ \hline
P1   & Cultural Exhibition Practitioner    & 6 years  \\
P2   & Industrial Design Student           & 4 years    \\
P3   & New Media Exhibition Practitioner   & 7 years    \\
P4   & Art \& Design Student               & 2 years    \\
P5   & Natural History Museum Researcher   & 9 years    \\
P6   & History Museum Researcher           & 24 years   \\
P7   & Municipal Museum Researcher         & 11 years   \\ \hline
\end{tabular}
}
\end{table}

\section{User Study}\label{A:eval}
\begin{table}[H]
\caption{Details of participants in the user study, including their professional backgrounds and museum or exhibition practice experience.}
  \Description{This table demonstrates the summary of participants interviewed in the user study.}
  \label{table:user participants}
\resizebox{0.48\textwidth}{!}{
\renewcommand\arraystretch{0.9}
\begin{tabular}{ccc}
\hline
ID   & Background                          & Domain Experience\\ \hline
U1   & Exhibition Practitioner    & 7 years  \\
U2   & Art \& Design Student           & 2 years    \\
U3   & Art \& Design Student and Exhibition Enthusiast  & 3 years    \\
U4   & Art \& Design Student               & 1 year    \\
U5   & Industrial Design Student   & 1 year   \\
U6   & Industrial Design Student           & 2 years   \\
U7   & Industrial Design Student         & 1 year   \\ 
U8   & Exhibition Designer   & 8 years    \\
U9   & Art \& Design Student               & 1 year    \\
U10   & Art \& Design Student   & 2 years    \\
U11  & Art \& Design Student           & 2 years   \\
U12   & Exhibition Designer         & 9 years   \\ \hline
\end{tabular}
}
\end{table}

\section{SiMUSation}\label{A:simusation:details}
\begin{table}[H]
\centering
\caption{Key objective metrics for evaluating visitor experience and exhibition performance.}
\Description{This table lists the measures used to assess visitor engagement, including their definitions and corresponding references.}
\label{tab:evaluation_metrics_formula}
\resizebox{0.49\textwidth}{!}{%
\renewcommand\arraystretch{1.5}
\begin{tabular}{lp{7.5cm}}
\hline
\textbf{Measure} & \textbf{Definition \& Formula} \\ 
\hline
Visitor Count (VC) & Total number of visitors. ($VC = \text{Count of Visitors}$) \\
Average Holding Time (AHT) & Average time (s) visitors spend engaging with an exhibit ($AHT = \frac{\sum \text{Time Spent}}{\text{VC}}$). \\
Attracting Power (AP) & Proportion of visitors who approach a specific exhibit ($AP = \frac{\text{Stopped Visitors}}{\text{VC}} \times 100\%$). \\
Sweep Rate Index (SRI) & Rate of movement through the exhibition (Area/Time). Lower is better for engagement ($SRI = \frac{\text{Area } (m^2)}{\text{Avg. Visit Time}}$). \\
Percentage of Diligent Visitors (\%DV) & Percentage of visitors stopping at $>50\%$ of exhibits ($\%DV = \frac{\text{Count of Visitors (>50\% Stops)}}{\text{Count of Visitors}} \times 100\%$). \\
Walkthrough Rate (WR) & Percentage of visitors not stopping at any exhibit ($WR = \frac{\text{Count of Visitors (0 Stops)}}{\text{VC}} \times 100\%$). \\
Stops at Exhibits & Number of exhibits visited per visitor ($AvgStops = \frac{\sum \text{Total Stops}}{\text{VC}}$). \\
Dwell Time (DT) & Total time spent in the exhibition or a specific section ($DT = \text{Exit Time} - \text{Entry Time}$). \\
\hline
\end{tabular}%
}
\end{table}

\end{document}